\documentclass[letterpaper]{article} 
\usepackage{aaai24}  
\usepackage{times}  
\usepackage{helvet}  
\usepackage{courier}  
\usepackage[hyphens]{url}  
\usepackage{graphicx} 
\usepackage{natbib}  
\usepackage{caption} 
\usepackage{multirow}
\usepackage{xcolor,colortbl}
\definecolor{mygreen}{RGB}{0,255,0}
\definecolor{mygrey}{RGB}{192,192,192}

\usepackage{amsmath}
\usepackage{array}
\usepackage{siunitx}
\usepackage{dcolumn}
\usepackage{algorithm}
\usepackage{algorithmic}
\usepackage{comment}
\usepackage{amsmath,mathrsfs}
\usepackage{amsfonts}
\usepackage{bbding}
\usepackage{amssymb}
\usepackage{caption}
\usepackage{subfigure}
\usepackage{booktabs} 
\usepackage{threeparttable}
\usepackage{url}
\usepackage{cancel}
\usepackage{tabulary}
\usepackage{multirow}

\usepackage{listings} 
\usepackage{xcolor}
\usepackage{bibentry}
\usepackage{makecell}

\newcommand{\mycomment}[1]{10}

\usepackage{newfloat}
\usepackage{listings}
\DeclareCaptionStyle{ruled}{labelfont=normalfont,labelsep=colon,strut=off} 
\floatstyle{ruled}
\newfloat{listing}{tb}{lst}{}
\floatname{listing}{Listing}
\newcolumntype{d}[1]{D{.}{.}{#1}}

\title{Another Way to the Top: Exploit Contextual Clustering in Learned Image Coding}
 
\author {
    Yichi~Zhang\textsuperscript{\rm 1,2},
    Zhihao~Duan\textsuperscript{\rm 2},
    Ming~Lu\textsuperscript{\rm 3},
    Dandan~Ding\textsuperscript{\rm 1}\thanks{Corresponding author},
    Fengqing~Zhu\textsuperscript{\rm 2},
    Zhan~Ma\textsuperscript{\rm 3}
}

\affiliations {
    \textsuperscript{\rm 1}Hangzhou Normal University, Hangzhou, Zhejiang, China \\
    \textsuperscript{\rm 2}Purdue University, West Lafayette, Indiana, U.S. \\
    \textsuperscript{\rm 3} Nanjing University, Nanjing, Jiangsu, China\\
    yichizhang@ieee.org, duan90@purdue.edu, minglu@nju.edu.cn, DandanDing@hznu.edu.cn, zhu0@purdue.edu, mazhan@nju.edu.cn
}

\begin{document}
\maketitle

\begin{abstract}
While convolution and self-attention are extensively used in learned image compression (LIC) for transform coding, this paper proposes an alternative called Contextual Clustering based LIC (CLIC) which primarily relies on clustering operations and local attention for correlation characterization and compact representation of an image. As seen, CLIC expands the receptive field into the entire image for {intra-cluster} feature aggregation. Afterward, features are reordered to their original spatial positions to pass through the local attention units for inter-cluster embedding. Additionally, we introduce the Guided Post-Quantization Filtering (GuidedPQF)  into CLIC, effectively mitigating the propagation and accumulation of quantization errors at the initial decoding stage. Extensive experiments demonstrate the superior performance of CLIC over state-of-the-art works: when optimized using MSE, it outperforms VVC by about 10\% BD-Rate in three widely-used benchmark datasets; when optimized using MS-SSIM, it saves more than 50\% BD-Rate over VVC. Our CLIC offers a new way to generate compact representations for image compression, which also provides a novel direction along the line of LIC development.

\end{abstract}

\section{Introduction} 
\label{sect:introduction}

Lossy image compression is one of the most fundamental issues in information theory and signal processing. Most existing methods for lossy image compression follow the scheme of \textit{transform coding}~\cite{tc}, where images are transformed to a latent space for de-correlation and energy compression, followed by quantization and entropy coding. Historically, traditional codecs such as JPEG~\cite{wallace1992jpeg}, BPG~\cite{sullivan2012hevcoverview}, and VVC~\cite{bross2021vvcoverview} have utilized simple linear transforms (e.g., the discrete cosine transform) to accomplish this goal.

\begin{figure}[t] 
    \centering 
    \includegraphics[width=0.9\linewidth]{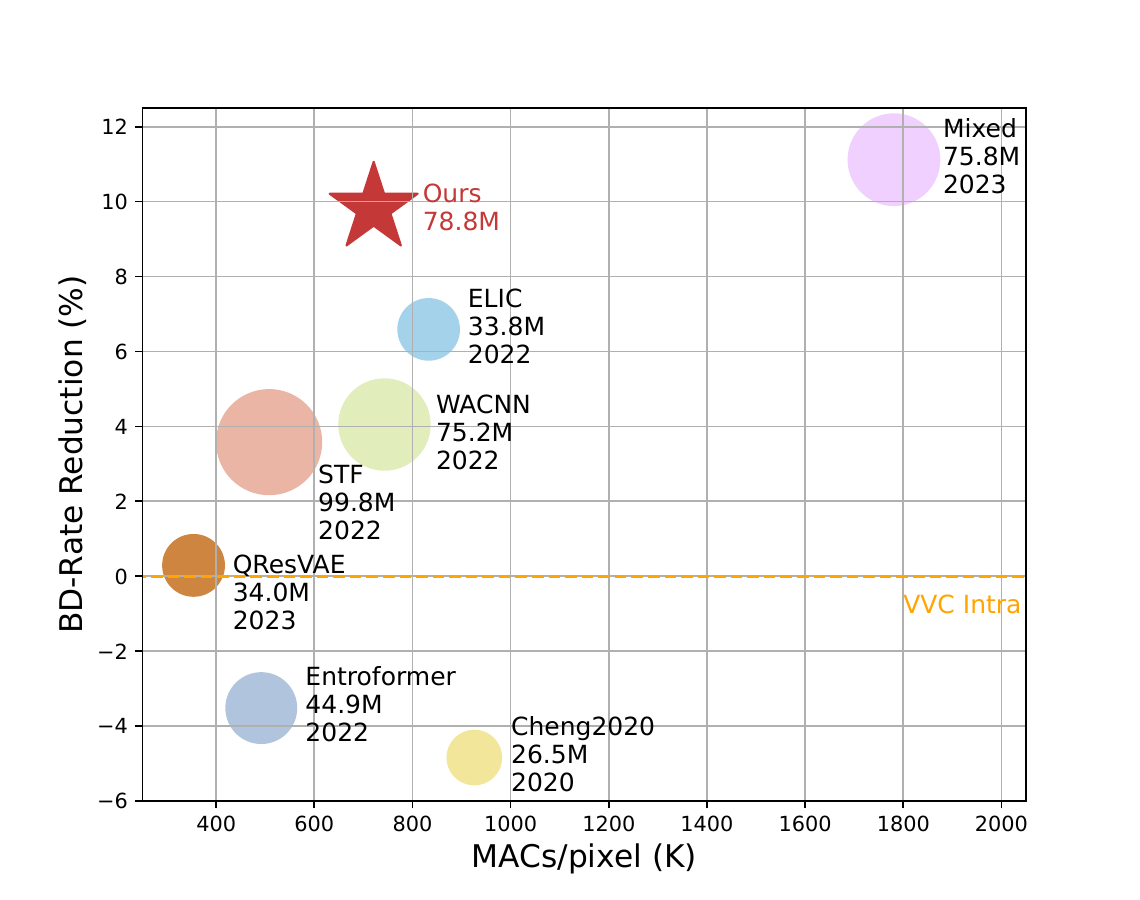}
    \caption{{Coding Performance versus End to End Complexity} across a variety of state-of-the-art LIC methods. MACs/pixel is calculated in the end-to-end manner, and BD-Rate is averaged in the Kodak dataset.} 
    \label{fig:front} 
\end{figure}

In recent years, learned image compression (LIC) methods have achieved superior performance over the traditional codecs~\cite{koyuncu2022contextformer,liu2023learned}. Central to the success of LIC is the utilization of non-linear transforms such as convolutional neural networks (CNNs) and Transformer modules. Early works stack standard convolutional layers for feature extraction and image reconstruction~\cite{balle2016end, balle2018variational}. Later, deformable convolutions~\cite{zhu2019deformable}, octave convolutions~\cite{chen2022two}, and asymmetric convolutions~\cite{AsymmetricConv} are developed to improve the standard convolutions. Recent works also introduce Transformers into CNNs~\cite{liu2023learned}. Although these convolution and attention-based methods consistently improve the rate-distortion (RD) performance of LIC, they often come at the expense of increased computational complexity, as depicted in Figure~\ref{fig:front}.

In contrast, this paper aims at improving the transform and quantization in the (non-linear) transform coding framework without increasing the overall computational complexity.
We first propose ``Contextual Clustering based LIC (CLIC)'', an alternative approach to convolution and Transformer-based methods, to exploit and characterize the spatial correlation within an image from a new perspective.
Specifically, inspired by~\cite{ma2023image}, we break the regular convolution operations on pixels, but reorganize all pixels in an image into several categories according to their similarities and apply Multilayer Perceptron (MLP) for {intra-cluster} correlation characterization. Furthermore, local attention units like spatial attention~\cite{woo2018cbam} and channel attention~\cite{hu2018squeeze} are embedded for {inter-cluster} exploration to augment the coding performance. In this way, the CLIC approach greatly reduces convolutions that require higher parameters and MACs by replacing them with simple linear and attention operations.

We also notice that the quantization operation following the transform step inevitably introduces quantization errors, adversely impacting the quality of reconstructions. To this end, filters are usually appended to decoded images for quality improvement~\cite{li2019deep,zhang2023content}. This, however, ignores the implicit propagation and accumulation of quantization error in the decoding process, which may limit the filtering performance~\cite{fu2023asymmetric}. To tackle this issue, we develop a Guided Post-Quantization Filtering method that compensates for the error at the beginning of decoding to prevent error propagation. It utilizes a set of coefficients supervised by the true errors to guide the decoder for filtering, enabling content-adaptive processing. By encoding these coefficients into the bitstream, the rate-distortion performance is improved with negligible complexity overhead.

Extensive experiments demonstrate the superior performance of our CLIC, as illustrated in Figure~\ref{fig:front}. Replacing the clustering operations with conventional convolutions in CLIC not only results in a slight decrease in coding performance but also increases computational complexity by 25\%. Overall, CLIC provides an alternative and promising solution for learned image compression beyond merely incremental performance improvement.

Our contributions are summarized as follows:
\begin{itemize}
    \item We propose the Contextual Clustering based approach, termed CLIC, mostly utilizing linear-based clustering and local attention to characterize pixel correlations in an image-level receptive field, outperforming convolution-based and window attention-based networks;
    \item We propose Guided Post-Quantization Filtering (GuidedPQF) to mitigate the propagation and accumulation of quantization errors while enabling content-adaptive processing at the initial decoding stage, improving rate-distortion performance;
    \item Our CLIC achieves around $10\%$ BD-Rate reduction over VVC on various image test sets, surpassing the performance of most existing approaches. Extensive experiments are conducted to verify our findings and demonstrate the effectiveness of CLIC.
\end{itemize}

\begin{figure*}[htbp] 
    \centering 
    \includegraphics[width=\linewidth]{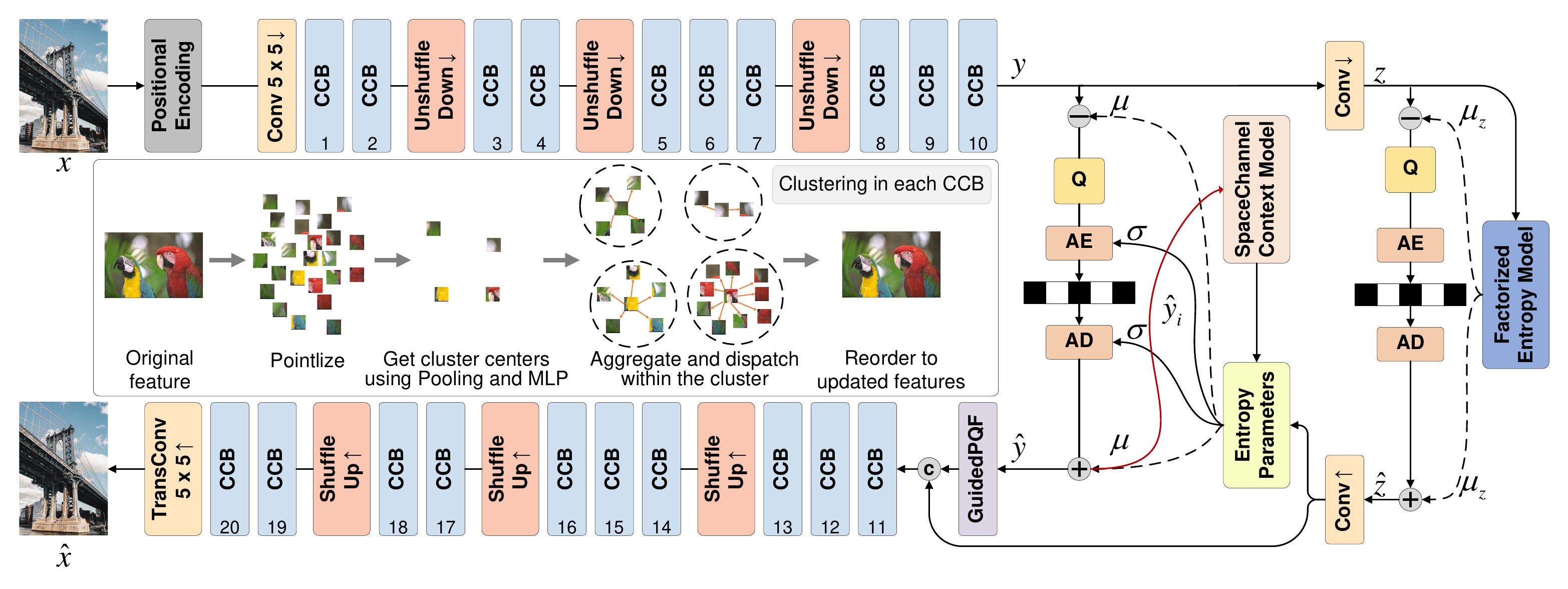}
    \caption{{The overall architecture of the proposed method.} $\downarrow$/$\uparrow$ indicates downsampling/upsampling operations. The hyper encoder and decoder each consist of five convolutional layers, succeeded by a GELU function, except for the final layer. The channel of $z$ is 192. Three hyper decoders are utilized for the mean, scale, and latent feature, respectively.} 
    \label{fig:Overview} 
\end{figure*}

\section{Related Work} 
\label{sect:related_work}

To achieve RD optimization, two crucial techniques, transform coding and context modeling, are used for the generation and the entropy coding of image latent features in LIC, respectively. We will next review these two techniques. 


\subsection{Transform Coding}

Recently, learning-based transforms~\cite{chadha2021deep,ma2023rate,zhang2022unified} have shone in both rules-based and learned-based image compression. 
Regarding the transforms in LIC, stacked convolution layers are widely used to transform an image into another latent space for sparse representation. For example, Ball{\'e}~{\it et al.}~\cite{balle2016end} applied stacked convolutions in the VAE, together with Generalized Divisive Normalization (GDN), Inverse GDN (IGDN), for the transform function.
In addition to convolutions, attention modules~\cite{cheng2020learned} were introduced for compact representation of images, offering the first LIC method that obtained comparable performance to VVC. Then, convolutions and non-local attention modules were jointly utilized to derive latent features and hyperpriors and achieved impressive results~\cite{chen2021end}. 

Moreover, the prevalence of self-attention-based Transformers has inspired many researchers to investigate the use of Transformers for nonlinear transforms. Lu~{\it et al.}~\cite{lu2022transformer} introduced neural transformation units, which combined a Swin Transformer Block and a convolutional layer for the compact representation of images. Zhu~{\it et al.}~\cite{zhu2022transformer} employed Swin Transformer while Zou~{\it et al.}~\cite{zou2022devil} utilized a symmetrical Transformer for LIC. 
More recently, Liu~{\it et al.}~\cite{liu2023learned} mixed the structures of convolutions and Swin Transformer, achieving state-of-the-art results.

\subsection{Context Modeling}

Context Model significantly affects coding efficiency. In rules-based codecs, context-adaptive variable-length coding (CAVLC)~\cite{moon2005efficient} and context-adaptive binary arithmetic coding (CABAC)~\cite{sze2012high} are widely used to reduce redundancy by exploring the statistical correlation across coding symbols. In LIC, context models are also devised with the same goal of coding symbols using the lowest bitrate. Typical methods include the autoregressive context model~\cite{minnen2018joint} and its variants
~\cite{qianGlobalRef,koyuncu2022contextformer,kim2022joint,qianentroformer}.

However, these methods require long decoding times due to the nature of serial processing, hindering their use in practice. To this end, parallel context modeling methods, which are more friendly to real applications, are developed. The checkerboard model~\cite{he2021checkerboard} is a typical tool, in which the anchor content is encoded independently while the non-anchor content is encoded at a lower cost depending on the anchor content priors. Later, a generalized checkerboard~\cite{lu2022high} and a dual spatial prior model~\cite{guo2023evc} are introduced.

In addition to exploiting spatial correlation in an image, numerous works have been devoted to exploring the correlation across channels. The channel conditional model was devised to divide channels into slices for parallel processing~\cite{minnen2020channel} and later it was combined with the autoregressive and hierarchical prior entropy model to form a cross-channel context model~\cite{ma2021cross}.
Recently, He~{\it et al.}~\cite{he2022elic} noticed the uneven distribution of information among channels and proposed a channel-wise model with uneven grouping, termed the space-channel context model (SCCTX). Due to the well-balanced coding efficiency and time complexity of SCCTX, we directly use SCCTX for context modeling in this work.

{\bf Remarks.} Previous studies have extensively demonstrated the outstanding capability of convolutions for the transform coding of LIC. However, the mountaintop embraces diverse paths that lead to it. While convolutions offer promising coding efficiency, we present another elegant way by leveraging contextual clustering in LIC, which achieves even higher performance than previous works.

\section{Proposed Method}
\label{sect:proposed_method}
\subsection{Overview}

\begin{figure}[htbp] 
\centering 
\subfigure[Downsampling]{\includegraphics[width=0.30\linewidth]{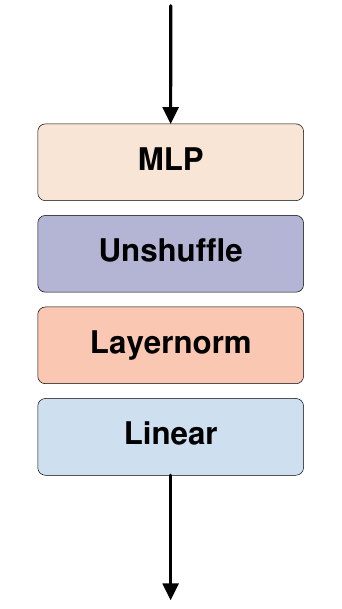}\label{subfig:UnshuffleDown}} 
\hspace{10mm}
\subfigure[Upsampling]{\raisebox{0.32cm}{\includegraphics[width=0.27\linewidth]{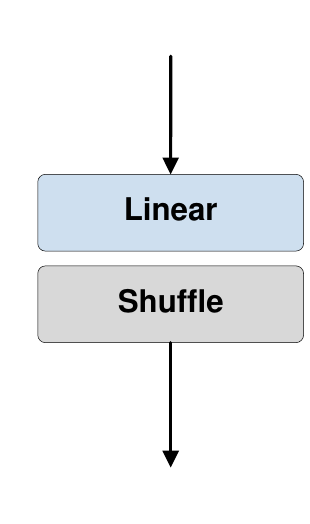}}\label{subfig:ShuffleUp}}
\caption{Strucuture of Down / Up sampling} 
\label{fig:updown} 
\end{figure}

An overview of the proposed CLIC is illustrated in Figure~\ref{fig:Overview}. Both the main analysis transform module and the main synthesis transform module comprise four stages, each containing one downsampling/upsampling operation and a set of Contextual Clustering Blocks (CCBs).
The downsampling operation stacks MLP, Unshuffle, Layernorm, and Linear layers, while the upsampling operation only goes through a linear layer and a shuffle layer, as shown in Figure ~\ref{fig:updown}. After downsampling/upsampling are stacked CCBs to exploit spatial correlations across pixels for a compact representation of the input image.

At the beginning of the analysis transform, we cascade the attribute features (RGB color) of each pixel with its position features (Cartesian coordinate system (X, Y) coordinates). In this way, an image $I \in \mathbb{R}^{H\times W\times 3}$ is transformed into $P\in \mathbb{R}^{n\times 5}$ points, $n = H \times W$, with each point consisting of its RGB attribute $(r,g,b)$ and position information $
(x,y)$ (for Positional Encoding).  For the context modeling, we directly combine the checkboard spatial model and uneven channel-wise to generate SCCTX, and the latent feature for reconstruction is also utilized to fully explore the side information. More details can refer to~\cite{he2022elic,hu2020coarse}.

At the beginning of the main synthesis transform, GuidedPQF is applied to compensate for the quantization errors in the latent feature domain. Afterward, features are progressively upsampled and decoded. In the end, the reconstructed image is produced through a 5$\times$5 transposed convolution.

In the next, we will detail the algorithms and structures of CCB and GuidedPQF.

\subsection{Contextual Clustering Block}
\begin{figure}[t] 
    \centering 
    \includegraphics[width=\linewidth]{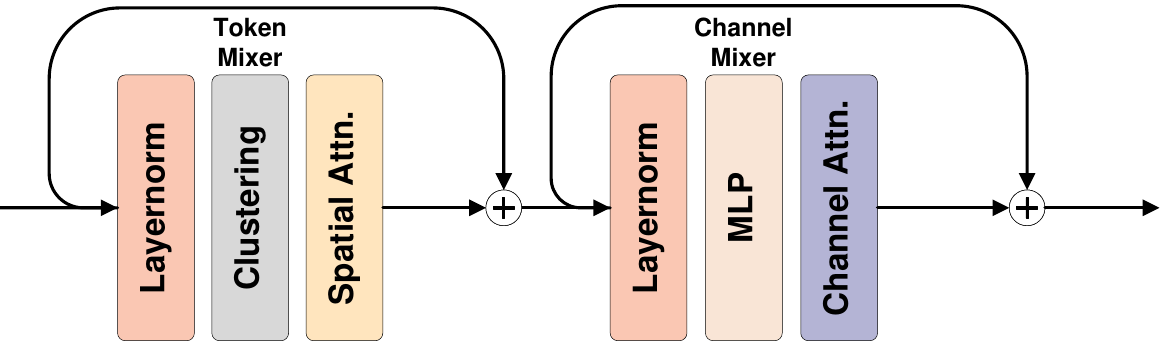}
    \caption{{The structure of CCB.} The MLP contains two linear layers and a GELU activation function.} 
    \label{fig:CCB} 
\end{figure}

As depicted in Figure~\ref{fig:CCB}, CCB is built on the structure of Metaformer~\cite{yu2022metaformer}, which consists of two layernorm~\cite{ba2016layer}, Token Mixer, Channel Mixer, and a residual connection. In our CLIC, the Token Mixer is implemented using the clustering approach~\cite{achanta2012slic,ma2023image}, followed by the spatial attention unit for feature enhancement. The Channel Mixer is realized using a simple MLP, followed by the channel attention unit. As observed and demonstrated in our experiments, clustering and attention units play crucial roles within each CCB.

{\bf Contextual Clustering.}
In contrast to convolutions that treat the image as a regular grid and operate in a fixed-size receptive field for processing, CCB considers the image as an unordered set of points (each point $P_i$ is a vector) and conducts feature extraction through clustering. {Each point contains its attribute features and position features.}

The clustering results greatly affect the performance of CCB. After extensive experiments, we resort to cosine similarity for clustering. The corresponding clustering algorithm is described as follows:
\begin{enumerate}
\item First, add the global context to all points by global average pooling:
\begin{equation}
    P_i = P_i + \gamma \cdot \frac{1}{n} \sum_{i=1}^n P_i,
\end{equation}
where $\gamma$ is a learnable parameter to scale the global context dynamically.
Then, all the pixel points are linearly transformed.
\item Get $c$ clustering centers by average pooling the feature map (i.e., all points $P$). In our experiments, we choose $c = 2 \times 2 = 4$ by default (we discuss the choice of $c$ in our ablation study). Then, a 2-layer MLP is used to predict the offset of each clustering center.
\item Calculate the cosine similarity matrix for all points with $c$ clustering centers, and the points most similar to each center are grouped.

\item {Within each cluster, points are aggregated based on their similarity to the clustering center.} Suppose we have $m$ points for a cluster. {We first linearly transform all the points to get their value vectors $v_i, \forall i \in \{1,2,...,m\}$, and their center point $c_v$ is obtained in the same way as in step 2.} The output feature $F$ for this cluster is computed by:
\begin{small}
\begin{equation}
    F=\frac{1}{1+m}\left(c_v+\sum_{i=1}^m \operatorname{sigmoid}\left(\alpha \cdot s_i +\beta\right) \cdot v_i\right),
\end{equation}
\end{small}

\noindent where $\alpha$ and $\beta$ are learnable parameters, and $s_i$ is the cosine similarity between the $i$-th point and the center. According to~\cite{ma2023image}, this gives better performance in a stable manner due to the non-negativity of its similarity. $c_v$ is used to emphasize the attribution of its class. {The normalization term $\frac{1}{1+m}$ controls the output feature magnitude, where we use $(1 + m)$ as the denominator to avoid division by zero when there is no point in a cluster (i.e., $m = 0$).}
\item The aggregated features are then adaptively assigned to each point in the cluster. For each point $P_i$, it is updated by $P_i \leftarrow P_i + \text{Linear}(s_i \cdot F)$. A linear layer is applied to match the dimension of $F$ and $P_i$.
\end{enumerate}

Moreover, we apply slightly different processing on odd-numbered and even-numbered CCBs. The odd-numbered CCBs exactly follow the above steps. For the even-numbered CCB, we divide features into two groups for separate processing in a checkboard manner. When one group passes through CCB, the other group is masked as zero.
{Such a checkerboard pattern enables the separation of features that have already been clustered together under one category in the odd-numbered CCBs, so as to generate new clustering results in the even-numbered CCBs at the same scale. As such, CLIC is able to derive various clustering features and more effectively utilize contextual information to improve coding efficiency.}

{\bf Reordering.} As seen, the clustering operation typically exploits point relationships within each cluster. Furthermore, after clustering, latent features obtained are reordered to the image shape before clustering according to their positions. Attention units are then applied for further processing.

\begin{figure}[h]
  \centering
\newcommand{\mywidth}{0.48}
   \includegraphics[width=\mywidth\linewidth]{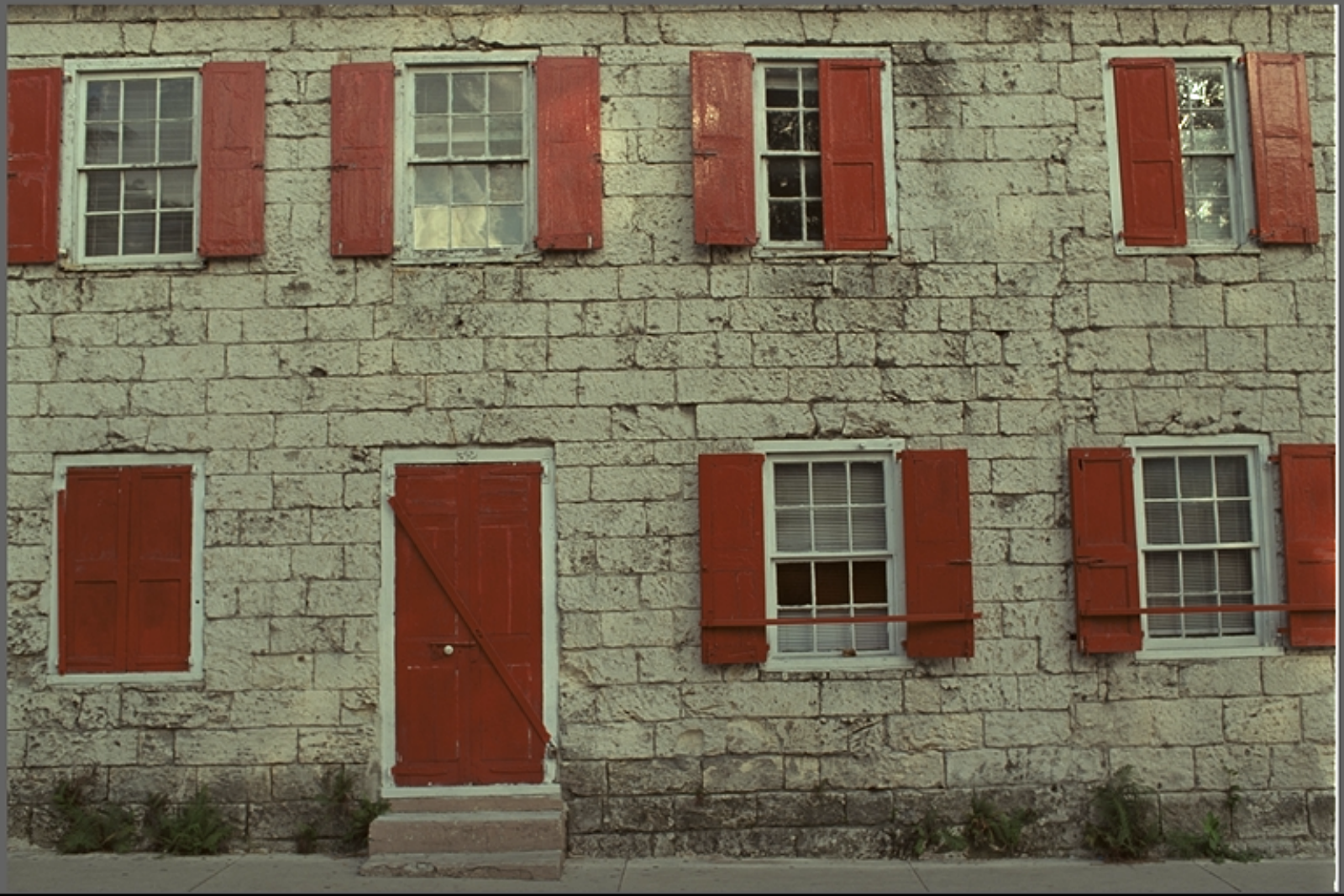}
      \includegraphics[width=\mywidth\linewidth]{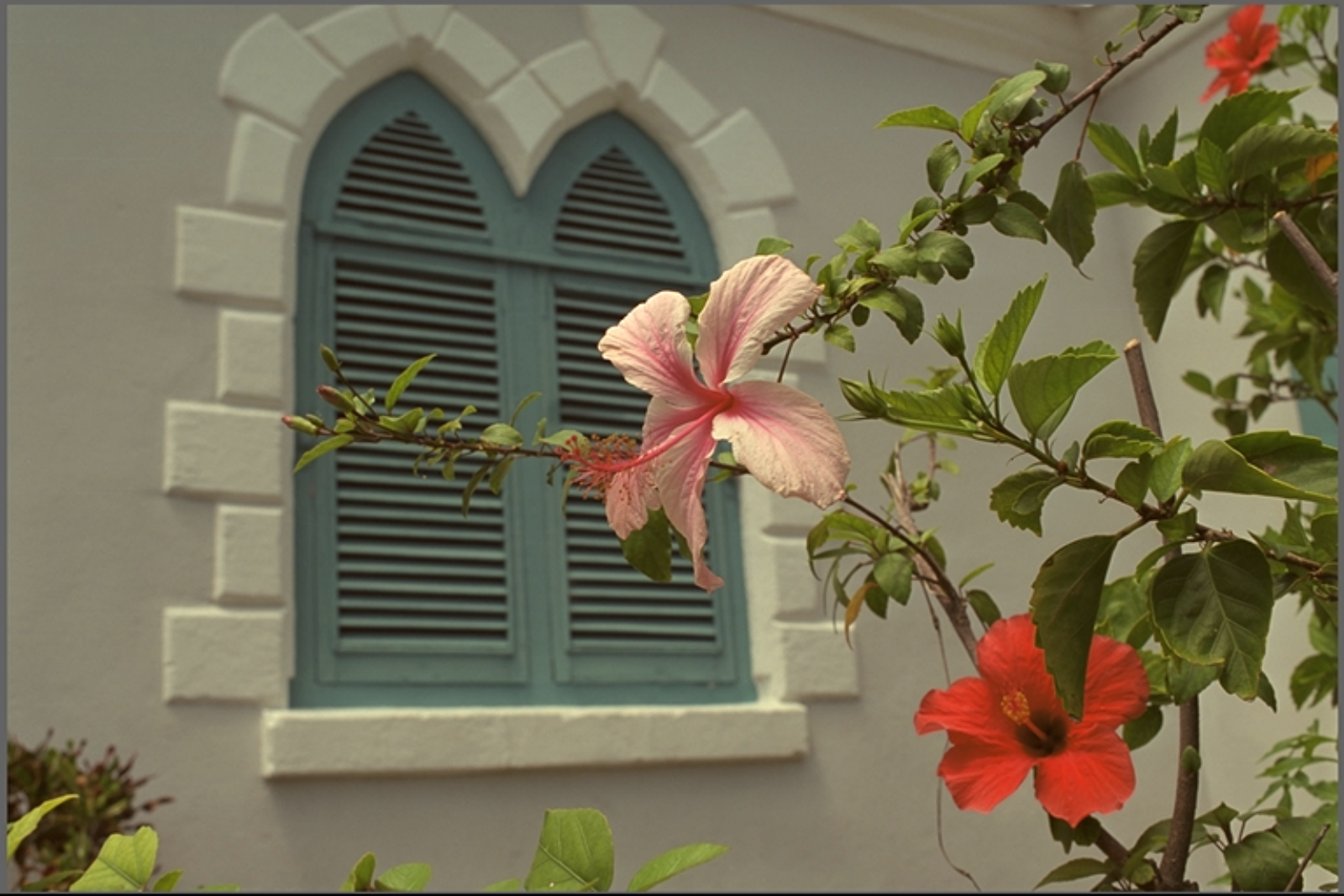} \\
      \includegraphics[width=\mywidth\linewidth]{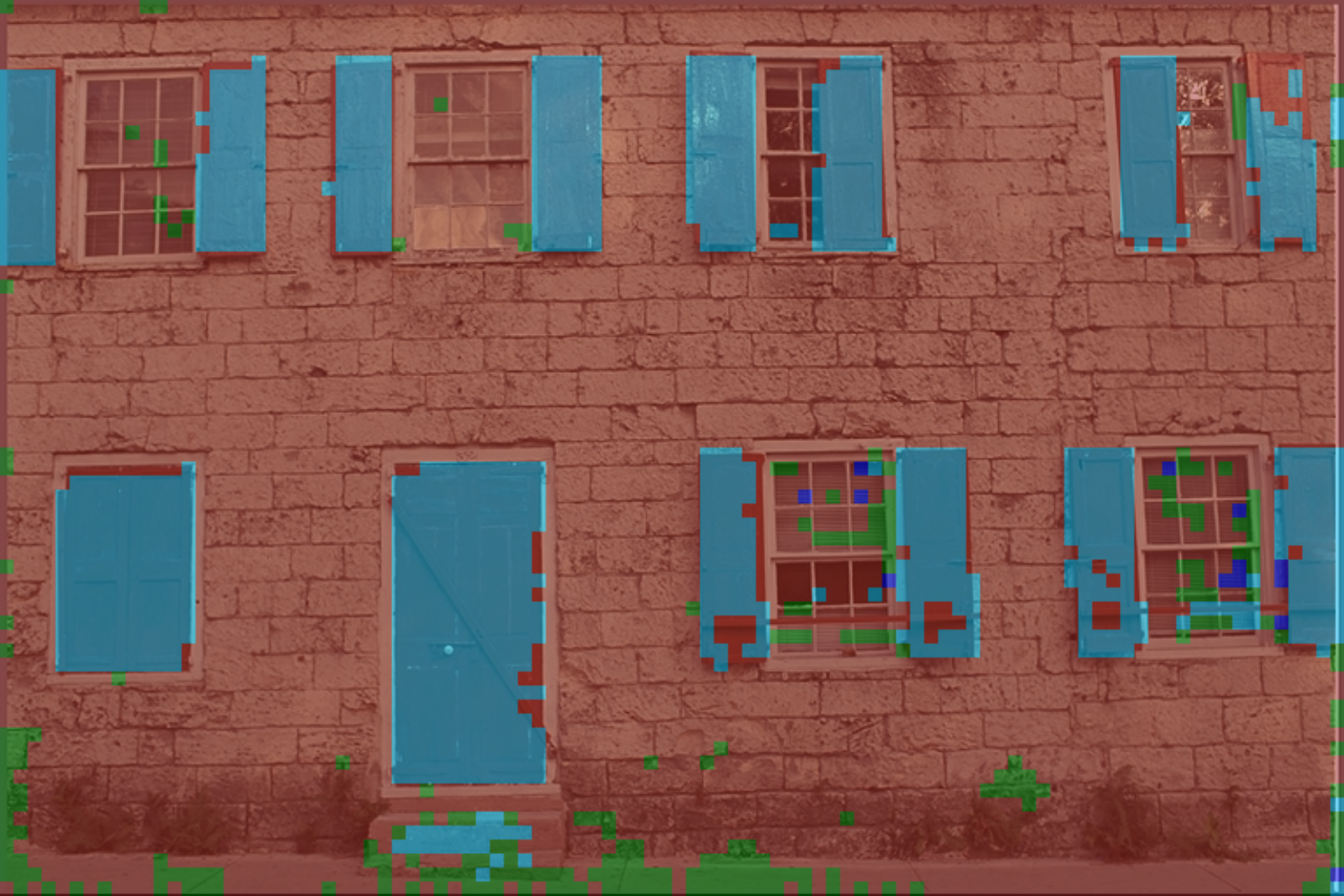}
      \includegraphics[width=\mywidth\linewidth]{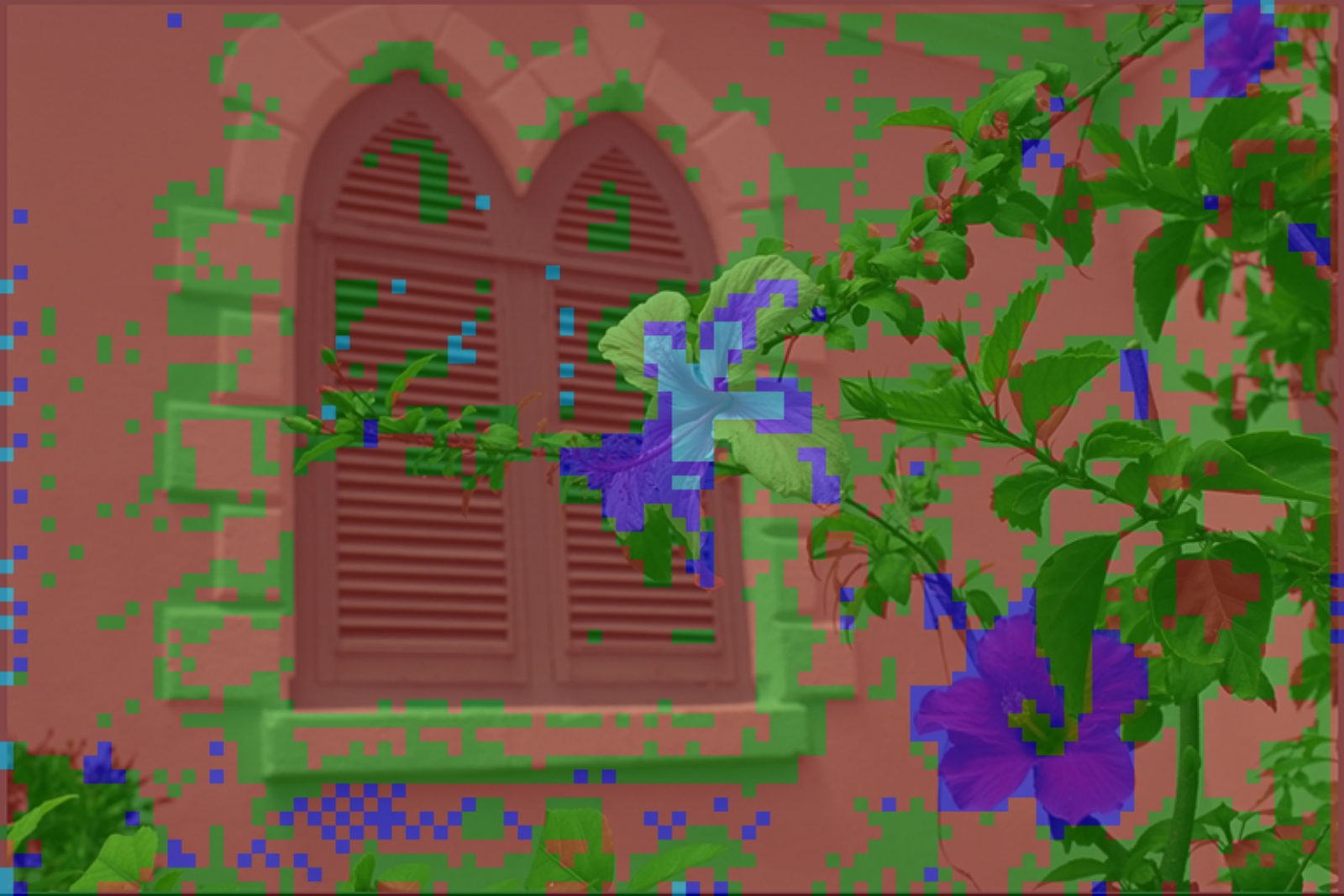}
  \caption{{Clustering results on Kodak.} The same color mask represents the same category. Clustering results are obtained from the final CCB.}
\label{fig:cluster_result}
\end{figure}

\begin{figure}[h] 
\centering 
\subfigure[]{\includegraphics[width=0.6\linewidth]{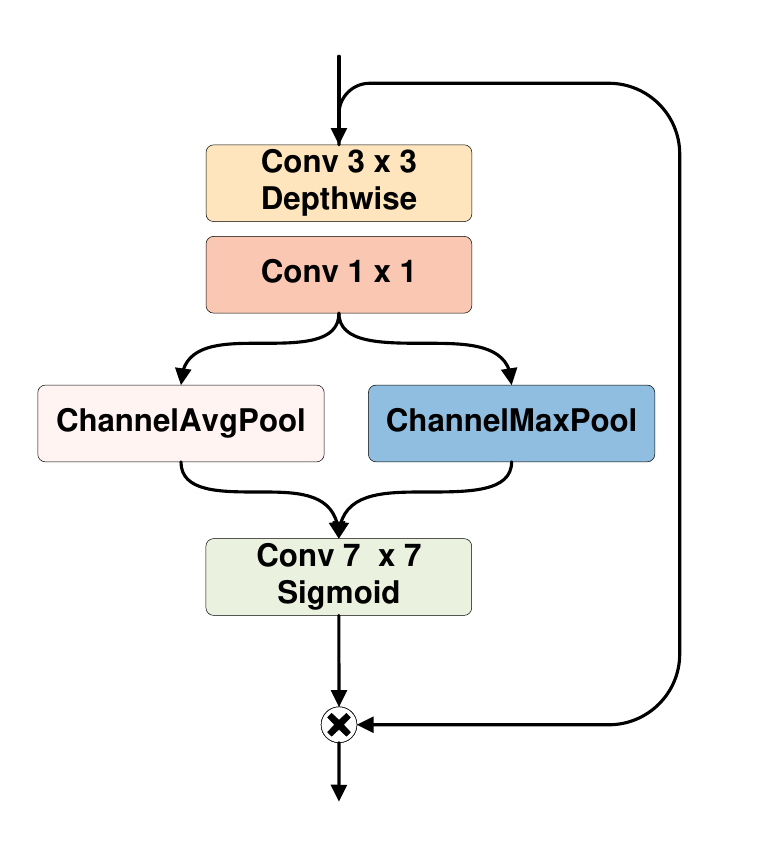}\label{subfig:Spa}}
\hspace{8mm}
\subfigure[]{\includegraphics[width=0.28\linewidth]{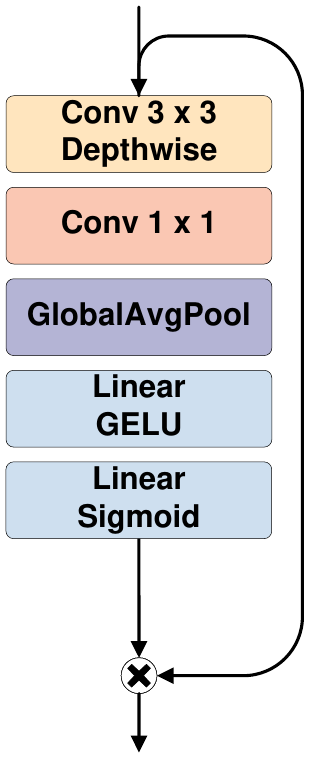}\label{subfig:Chan}} 
\caption{{Modular components:} (a) Spatial Attention; (b) Channel Attention.} 
\label{fig:modular} 
\end{figure}

{\bf Attention Enhancement.}
The contextual clustering approach primarily focuses on intra-cluster interactions, which limits its local neighborhood correlations. To this end, we further augment the attention mechanism after contextual clustering to exploit the inter-cluster correlations.  As shown in Figure~\ref{fig:modular}, two local attention units, including Spatial Attention (SA) and Channel Attention (CA), are applied in each CCB following Token Mixer and Channel Mixer, respectively. In this way, both intra-cluster and inter-cluster correlations are included in CCB for efficient representation.

{\bf Remarks.} CCB systematically rearranges all points in an image, enabling global correlation perception and aggregation in a predefined clustering manner, as the clustering results visualized in Figure~\ref{fig:cluster_result}.
Unlike conventional Transformers that rely on computationally intensive matrix-based operations, the clustering algorithm only requires linear operations, potentially offering practical advantages on specific platforms. The subsequent local attention units further strengthen the valuable features across clusters. As a result, CCB and local attention units collaboratively contribute to the compact image representation.

\subsection{Guided Post-Quantization Filtering}
All features denoted as $y$ in Figure~\ref{fig:Overview} undergo quantization for entropy coding, which inevitably introduces quantization errors. Such errors will be propagated and accumulated from scale to scale during the decoding process, severely impairing the reconstructed quality~\cite{fu2023asymmetric}. To address this issue, we propose preventing such error propagation from the start, i.e., a GuidedPQF is applied on the dequantized $\hat{y}$ to compensate for the quantization errors.

\begin{figure}[h] 
    \centering 
    \includegraphics[width=0.758\linewidth]{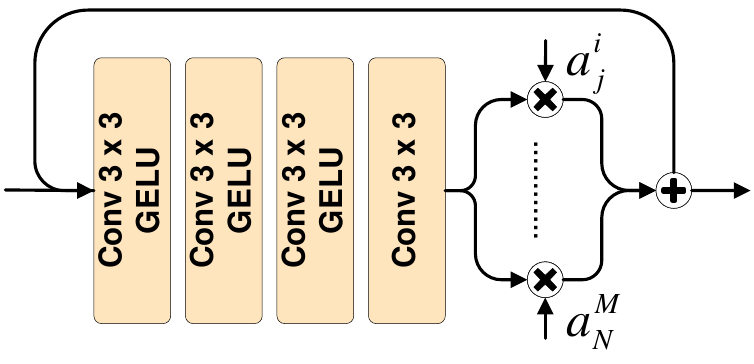}
    \caption{{The structure of GuidedPQF.}} 
    \label{fig:pqf} 
\end{figure}

To this end, the goal of GuidedPQF is to minimize the distance between the estimated and true quantization errors:
\begin{equation} \label{eq:pqf_1}
{e}= \|\tilde{\epsilon}-{\epsilon}\|^2= \|F(\hat{y} ; \theta)-{\epsilon}\|^2,
\end{equation}
where $\tilde{\epsilon} = \tilde{y} - \hat{y} = F(\hat{y}; \theta)$ represents the the estimated quantization error output from  GuidedPQF $F(\cdot)$. ${\epsilon} = {y} - \hat{y}$ denotes the true quantization error.

In contrast to previous studies that directly employ a neural model to realize the filter $F(\cdot)$, GuidedPQF utilizes a set of coefficients from the encoder to guide the filter for content-adaptive processing. As illustrated in Figure~\ref{fig:pqf}, given the dequantized features $\hat{y}=\{\hat{y}_i\}_{i=1}^{M}$, GuidedPQF will generate $N$ filtering candidates $\mathbf C_1^i, \mathbf C_2^i, \cdots, \mathbf C_{N}^i$ for each $\hat{y}_i$ and weigh them to derive $\tilde{\epsilon} = \{ \tilde{\epsilon}_i \}_{i=1}^{M} $ and then ${\tilde{y} = \{ \tilde{y}_i\}_{i=1}^{M}}$: 
\begin{equation} \label{eq:pqf_2}
 \tilde{y}_i = \tilde{\epsilon}_i 
 + \hat{y}_i = a_1^i \mathbf{C}_1^i + a_2^i \mathbf{C}_2^i + \cdots +a_{N}^i \mathbf{C}_{N}^i + \hat y_i,
\end{equation}
where $a_j^i$, $j\in \{ 1, 2, ..., N \}$ are weighting coefficients that can be signaled in bitstream. 

Based on Eq.~\eqref{eq:pqf_1} and Eq.~\eqref{eq:pqf_2}, given ${\epsilon} = \{{\epsilon}_i \}_{i=1}^{M} $, $a_j^i$ can be obtained by least square optimization~\cite{ding2023neural}:
\begin{equation} \label{eq:pqf_3}
\left[a_1^i, a_2^i, \cdots, a_{N}^i\right]^{\mathrm{T}}=\left(\mathbf{C}^{i\mathrm{T}} \mathbf{C}^i\right)^{-1} \mathbf{C}^{i \mathrm{T}} \epsilon_i,
\end{equation}
where $\mathbf{C}^i = [\mathbf{C}_1^i, \mathbf{C}_2^i, \cdots , \mathbf{C}_N^i]$ stacks vectorized $\mathbf {C}_j^i$s.

Substituting Eq.~\eqref{eq:pqf_3} into Eq.~\eqref{eq:pqf_1}, we can derive the reconstruction error $e_i$:
\begin{equation}
e_i  =\|\hat{\epsilon}_i-\epsilon_i\|^2 =|\epsilon_i|^2-\epsilon_i^{\mathrm{T}} \mathbf{C}^i\left(\mathbf{C}^{i \mathrm{T}} \mathbf{C}^i\right)^{-1} \mathbf{C}^{i \mathrm{T}} \epsilon_i.
\end{equation}

As a result, for $\hat{y}$ with $M$ channels, the objective function of our GuidedPQF can be described as:
\begin{equation}
\mathcal{{L}_{PQF}} = \sum_{i=1}^M\left\{\cancel{{\left|\epsilon_i\right|^2}} -\epsilon_i^{\mathrm{T} }\mathbf{C}^i\left(\mathbf{C}^{i \mathrm{T}} \mathbf{C}^i\right)^{-1} \mathbf{C}^{i \mathrm{T}} \epsilon_i\right\}.
\end{equation}

Since $\left|\epsilon_i\right|^2$ is a constant for each $\hat{y}_i$, we can directly remove it.
{
Finally, the overall loss function is a summation of $\mathcal{{L}_{PQF}}$ and the common rate-distortion loss function:
\begin{equation}
\mathcal{L}= \mathcal{R} + \lambda \cdot \mathcal{D} + \lambda_1 \cdot \mathcal{L_{PQF}},
\end{equation}
where the rate $\mathcal{R}$ and distortion $\mathcal{D}$ are computed following the standard practice in Hyperprior models~\cite{balle2018variational}. $\lambda_1$ is weighting factors adjusting magnitude orders and is set to 1 in our method.
}
As the weighting coefficients $a_j^i$s are obtained from the true error ${\epsilon}$, they have to be passed to the decoder by consuming certain bitrates. We use fix-length coding (four-bits representation) to encode $a_j^i$s and set $N=2$ to limit their bitrate increase.

\section{Experimental Results} 
\label{sect:experimental_results}

\subsection{Experimental Settings}
\textbf{Training.} We use Flicker2W~\cite{liu2020flicker2w} and LIU4K~\cite{liu2020comprehensiveLIU4k} as our training sets. We first scaled the longer side of the images in LIU4K to 2000 pixels with the same aspect ratio, and then randomly cropped $256\times256$ patches for training. Following Liu~{\it et al.}~\cite{liu2020flicker2w}, 99\% of the images were used for training, and the remaining 1\% were used for validation.  Following the settings
of CompressAI~\cite{begaint2020compressai}, we set $\lambda \in$ \{18, 35, 67, 130, 250, 483\} $\times 10^{-4}$ for MSE optimized model and $\lambda \in$ \{2.40, 4.58, 8.73, 16.64, 31.73, 60.50\} for MS-SSIM optimized model. We trained each model with Adam optimizer~\cite{kingma2014adam} with $\beta_1 = 0.9$, $\beta_2 = 0.999$. Each model was trained for 300 epochs with a batch size of 8 and an initial learning rate of $1e^{-4}$. We used the ReduceLRonPlateau lr scheduler with a patience of 5 and a factor of 0.5. The first 150 epochs are called Stage 1 and the next 150 epochs are called Stage 2. When switching from Stage 1 to Stage 2, the learning rate is newly set to $1e^{-4}$. We use mixed quantizers to train the channel-conditional models. We use universal quantization (U-Q)~\cite{choi2019variable} to generate U-Q${(y)}$ for the context model and use differentiable soft quantization (DS-Q)~\cite{gong2019differentiable} to generate DS-Q$(y)$ as input to the decoder. Also, following previous work~\cite{minnen2018joint,minnen2020channel,he2022elic} and community discussions\footnote{\url{https://groups.google.com/g/tensorflow-compression/c/LQtTAo6l26U/m/mxP-VWPdAgAJ}}, we do not encode $\lceil y \rfloor$ as a bitstream, instead, we encode each $\lceil y-\mu \rfloor$. When reverting the encoded symbols, we revert them to $\lceil y-\mu \rfloor + \mu$, which enables the single Gaussian entropy model to yield better results.

All training was performed on a computer with an RTX4090 GPU, i9-13900K CPU, and 64G RAM. Ablation experiments were performed on a computer with an RTX3090 GPU, i7-9700K CPU, and 64G RAM.

\textbf{Testing.} Three widely-used benchmark datasets, including Kodak\footnote{\url{https://r0k.us/graphics/kodak/}}, Tecnick\footnote{\url{https://tecnick.com/?aiocp\%20dp=testimages}}, and CLIC 2022\footnote{\url{http://compression.cc/}}, are used to evaluate the performance of the proposed method.

\subsection{Quantitative Results}

\begin{figure*}[t] 
\newcommand{\mywidth}{0.33}
\centering 
\subfigure[Kodak (MSE)]{\includegraphics[width=\mywidth\linewidth]{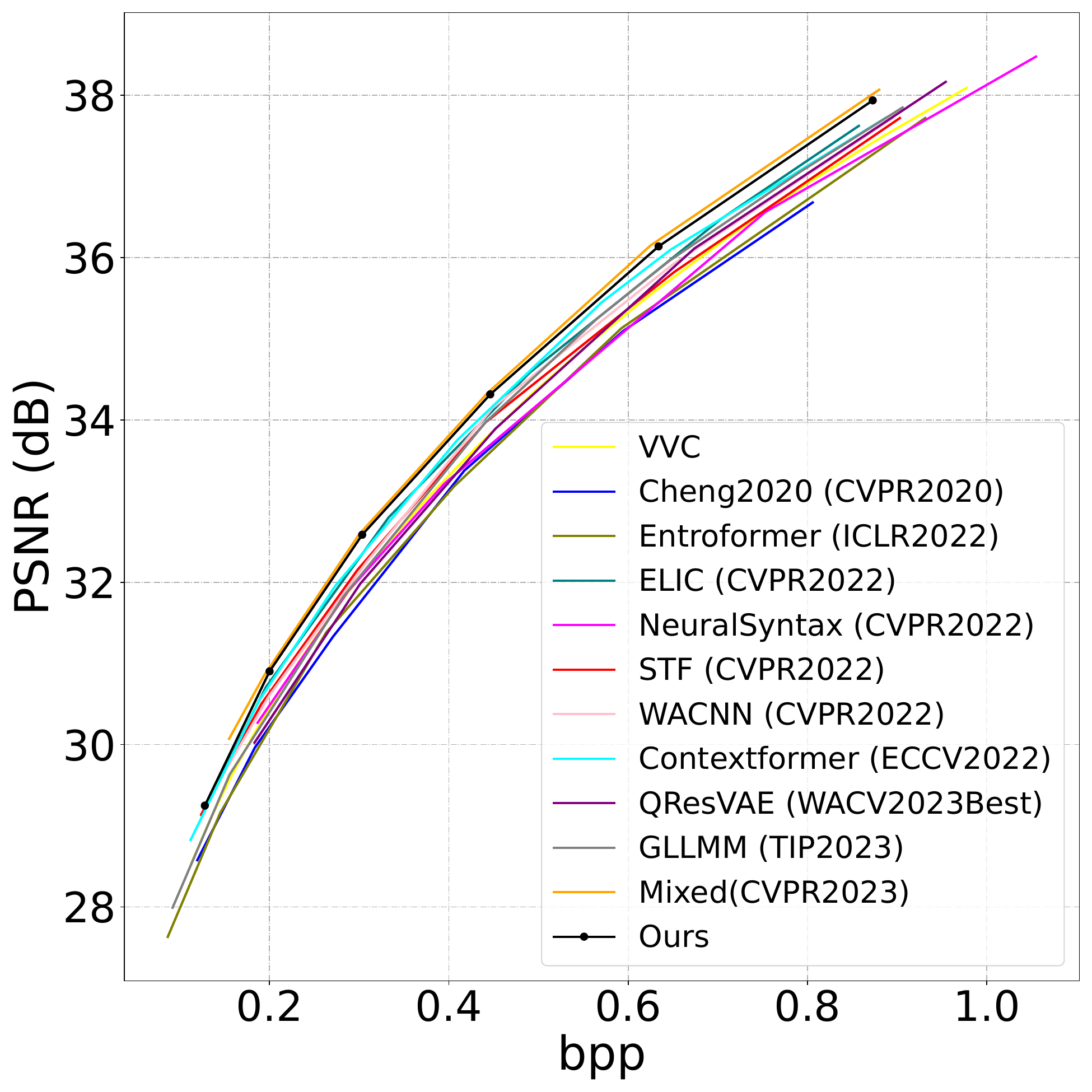}\label{subfig:kodak}} 
\subfigure[Tecnick 1200x1200 (MSE)]{\includegraphics[width=\mywidth\linewidth]{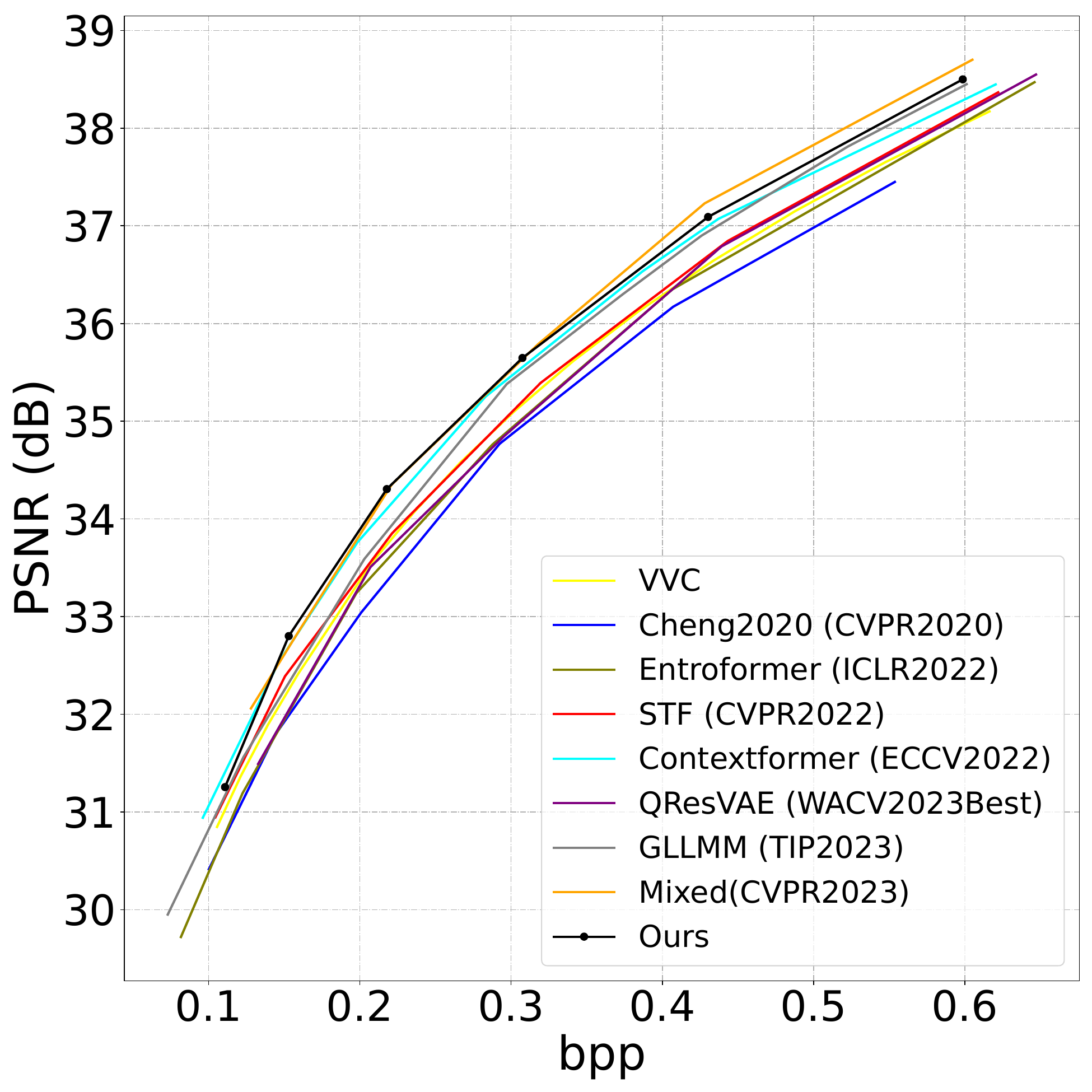}\label{subfig:Tecnick}}
\subfigure[CLIC 2022 (MSE)]{\includegraphics[width=\mywidth\linewidth]{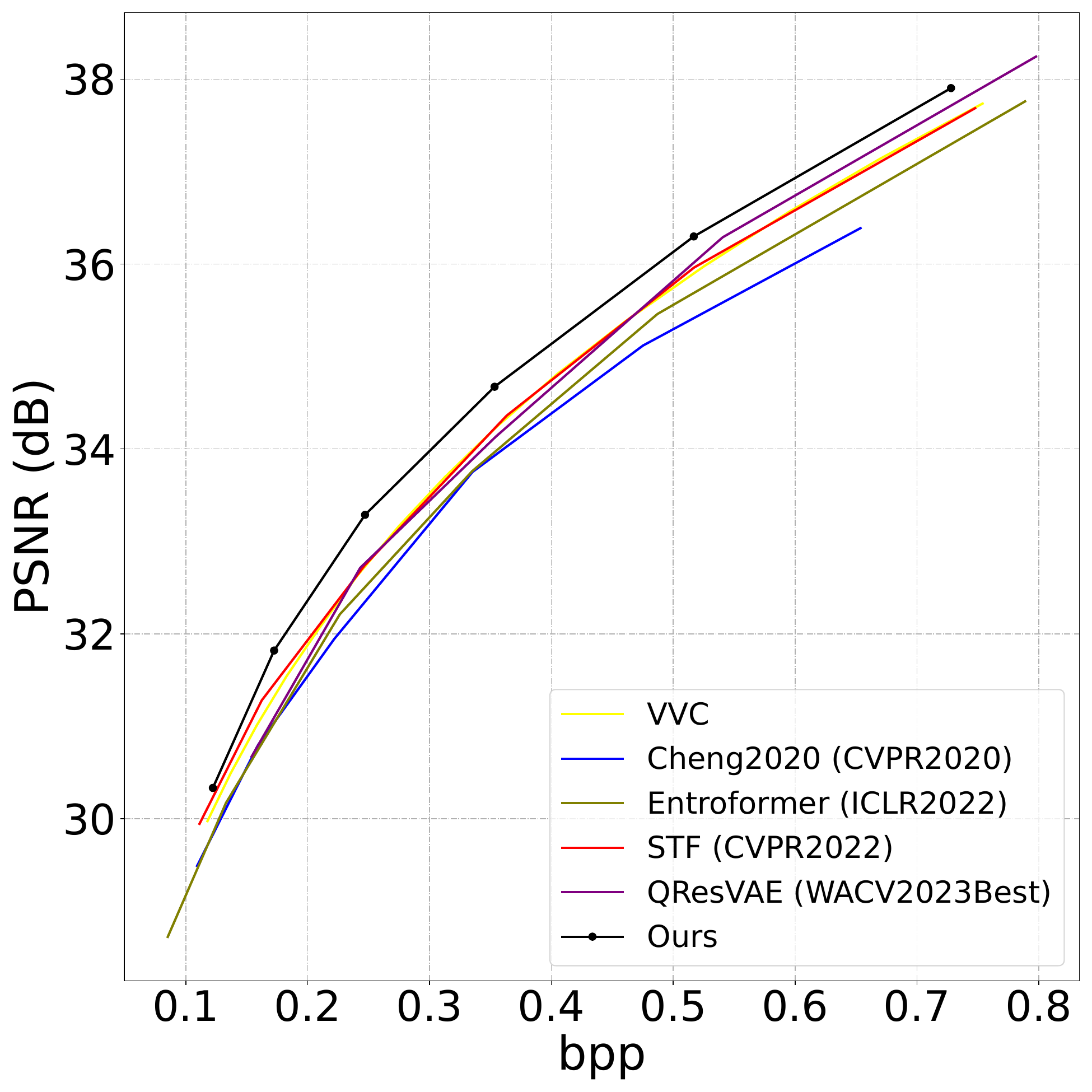}\label{subfig:CLIC}}
\caption{{RD curves of various methods.} (a), (b), and (c) present MSE-optimized results. {\it Please zoom in for more details}.} 
\label{fig:mse_result} 
\end{figure*}

\begin{table}[htbp]
  \centering
  \fontsize{15pt}{17pt}\selectfont
  \resizebox{\linewidth}{!}{
    \begin{tabular}{c|c|c|c|c}\hline\hline
    \multirow{2}[3]{*}{Method} & \multicolumn{3}{c|}{MSE} & \multicolumn{1}{l}{MS-SSIM} \\
\cline{2-5}          & Kodak & Tecnick & \multicolumn{1}{c|}{CLIC2022} & \multicolumn{1}{c}{Kodak} \\\hline
    Cheng2020 & 4.84\% & 6.15\% & 8.29\% & -44.00\% \\
    Entroformer & 3.52\% & 2.04\% & 4.72\% & -43.73\% \\
    NeuralSyntax & 0.22\% & -     & {-} & {-} \\
    QResVAE & -0.29\% & 1.02\% & {{-0.13\%}} & {-} \\
    GLLMM & -2.64\% & -5.66\% & {-} & -48.32\% \\
    STF   & -3.58\% & -2.54\% & {{-1.62\%}} & -48.77\% \\
    WACNN & -4.05\% & -     & {-} & {{-48.86\%}} \\
    ELIC  & -6.59\% & -     & {-} & -45.09\% \\
    Contextformer & {{-6.92\%}} & {{-9.47\%}} & {-} & -46.66\% \\
    Mixed &{{-11.12\%}}&{{-11.79\%}}& - &{{-49.68\%}} \\\hline
    Ours  & {\textbf{-9.83\%}} & {\textbf{-11.00\%}} & {\textbf{-10.05\%}} & {\textbf{-51.71\%}} \\\hline\hline
    \end{tabular}}\selectfont%
    \caption{Average BD-Rate (\%) reduction against VVC anchor in different datasets}
  \label{tab:bdrate}%
\end{table}%

We compare our proposed method with prevalent learning-based image compression models including Cheng2020~\cite{cheng2020learned}, Entroformer~\cite{qianentroformer}, NeuralSyntax~\cite{wang2022neural}, QResVAE~\cite{duan2023lossy}, GLLMM~\cite{fu2023learned}, STF~\cite{zou2022devil}, WACNN~\cite{zou2022devil}, ELIC~\cite{he2022elic}, Contextformer~\cite{koyuncu2022contextformer}, and Mixed~\cite{liu2023learned} and rules-based VVC~\cite{bross2021vvcoverview}. We use VVC reference software VTM-18.0 under All Intra configuration as the anchor to calculate BD-Rate. RD points of ELIC, Contextformer, and Mixed are digitized from the figures in their original papers or websites since they are not open-source.

Table~\ref{tab:bdrate} reports the BD-Rate reduction of each method against the VVC anchor on three datasets. Our proposed method achieves near-best performance on each dataset, comparable to Mixed, on average 9.83\% on Kodak, 11.00\% on Tecnick, and 10.05\% on CLIC 2022. Figure~\ref{fig:mse_result} further plots RD curves of all methods. Moreover, our method consistently maintains around 10\% BD-Rate gain over VVC on the other two datasets, showcasing its strong generalization on various content and resolutions. 

{Our CLIC gains slightly lower than Mixed when optimized with MSE. This occurs mainly because: 1) our training dataset has only 21,600 images while Mixed's has 300,000 images; 2) Mixed stacks massive Convolutions and Transformers at the expense of highly intensive complexity, which is 2.47$\times$ of our MACs.}

In addition, the coding performance of MS-SSIM optimized models is presented in Table ~\ref{tab:bdrate}. Our method achieves the best results and is the only one that achieves over 50\% (i.e., 51.71\%) BD-Rate reduction among all methods. The second best method obtains 49.68\% BD-Rate gains, which is inferior to ours by 2.03\% BD-Rate.


\newcommand{\myfontsize}{\footnotesize}
\newcommand{\myspace}{0.3}
\begin{figure}[t]
  \centering 
  \begin{minipage}[b]{1\linewidth} 
  \centering
   \subfigure[Original]{
    \begin{minipage}[b]{\myspace\linewidth} 
      \centering
      \includegraphics[width=\linewidth]{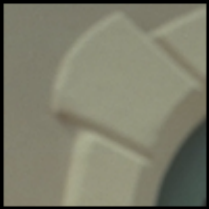}\\
      \begin{center} {\myfontsize PSNR / BPP} \end{center}
      \includegraphics[width=\linewidth]{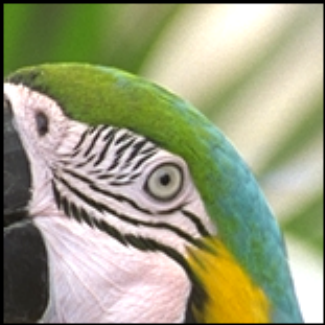}
      \begin{center} {\myfontsize PSNR / BPP} \end{center}
     
    \end{minipage}
  }
    \subfigure[VTM-18]{
    \begin{minipage}[b]{\myspace\linewidth} 
      \centering
      \includegraphics[width=\linewidth]{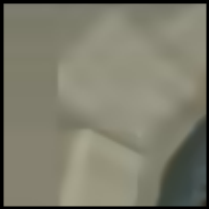} \\ 
      \begin{center} {\myfontsize 30.2338 / 0.1081} \end{center}
      \includegraphics[width=\linewidth]{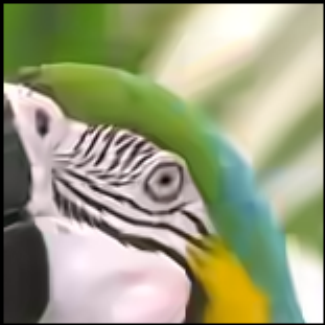}
      \begin{center} {\myfontsize 31.6807 / 0.0635} \end{center}
    \end{minipage}
  }\subfigure[Ours]{
    \begin{minipage}[b]{\myspace\linewidth} 
      \centering
      \includegraphics[width=\textwidth]{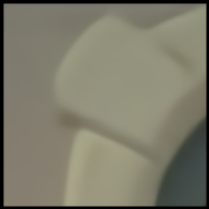} \\ 
      \begin{center} {\myfontsize {\textbf{31.3706 / 0.1068}}} \end{center} 
      \includegraphics[width=\linewidth]{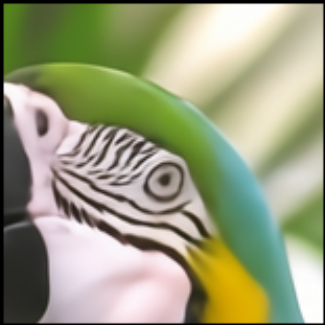}
      \begin{center} {\myfontsize {\textbf{32.7820 / 0.0659}}} \end{center}
     
    \end{minipage}
  }
   \end{minipage}
  \caption{{Visualization of the reconstructed images from the Kodak dataset.} Bold indicates the best performance.}
  \label{fig:Visual_compare}
\end{figure}

\begin{table}[t]
  \centering
  
\resizebox{\linewidth}{!}   {
  \begin{threeparttable}
  
    \begin{tabular}{c|c|c|c|c}\hline\hline
    \multirow{2}[0]{*}{Method} & \multirow{2}[0]{*}{Params} & {MACs} & \multirow{2}[0]{*}{Enc.} & \multirow{2}[0]{*}{Dec.} \\
          &       & (per pixel) &       &  \\\hline
   Cheng2020 & 26.5M & 926k  & 1.9855s & 4.0475s \\
    Entroformer & 44.9M & 492k  & -     & - \\
    NeuralSyntax & 52.9M & 7980k & -     & - \\
    QResVAE & 34.0M & 354k  & 0.1450s & 0.0551s \\
    GLLMM & -     & -     & -     & - \\
    STF   & 99.8M & 508k  & 0.1296s & 0.1242s \\
    WACNN & 75.2M & 743k  & 0.1148s & 0.1148s \\
    ELIC$^*$  & 33.8M & 833k  & 0.1716s & 0.0830s \\
    Contextformer$^*$ & -     & -     &  40s     & 44s \\
    Mixed & 75.8M     & 1781k     &  0.1405s
    & 0.1300 s \\\hline
    Ours  & 78.8M & 721k  & 0.1788s & 0.1012s \\\hline
    \end{tabular}%
    
  \begin{tablenotes}
  \item {\bf Test Conditions}: Intel i9-13900K CPU, Nvidia 4090 GPU, Windows 10. The enc./dec. time is averaged over all 24 images in Kodak, including entropy enc./dec. time. 
  \item $^*$: We reproduced ELIC~\cite{he2022elic} to calculate the runtime. The Enc. \& Dec. time of Contextformer is picked from~\cite{koyuncu2022contextformer}, which was tested on an NVIDIA Titan RTX GPU, i9-10980XE CPU.
   \item  MACs (per pixel) is calculated in an end to end manner.
  \end{tablenotes}
    \end{threeparttable}}
    \caption{Computational complexity compared with SOTAs}
     \label{tab:Complexity}%
\end{table}%

\subsection{Qualitative Visualization}

Figure~\ref{fig:Visual_compare} compares the qualitative results of our proposed CLIC and VVC. Here, the value of $\lambda$ is set as 0.0018 in CLIC (corresponding to VTM-18.0 QP 42). As observed, CLIC yields more visually pleasing reconstructed images, exhibiting clearer textures and less noise.

\subsection{Complexity}
Table~\ref{tab:Complexity} measures the computational complexity of each method using the number of parameters, MACs per pixel, encoding time (Enc.), and decoding time (Dec.). Our CLIC uses almost the same Context model as ELIC. Even though our number of parameters is higher than ELIC (78.8M vs. 33.8M), the computational complexity of MACs per pixel is significantly lower than ELIC (721K vs. 833K) due to the efficiency of the clustering method. Our encoding and decoding time is also comparable with ELIC and slightly higher than other channel-conditional methods. Overall, our CLIC outperforms ELIC by a large margin, 9.83\% vs. 6.59\%, with almost the same complexity. In addition, Our MACs are only 40\% of Mixed with a similar number of parameters (75.8M vs. 78.8M) and performance (1.29\% BD-Rate lower on Kodak, 0.79\% BD-Rate lower on Tecnick, 2.03\% BD-Rate higher on MS-SSIM optimized Kodak).

\begin{figure*}[t] 
\centering 
\newcommand{\mywidth}{0.33}
\newcommand{\myhspace}{-1mm}
\subfigure[CCB]{\includegraphics[width=\mywidth\linewidth]{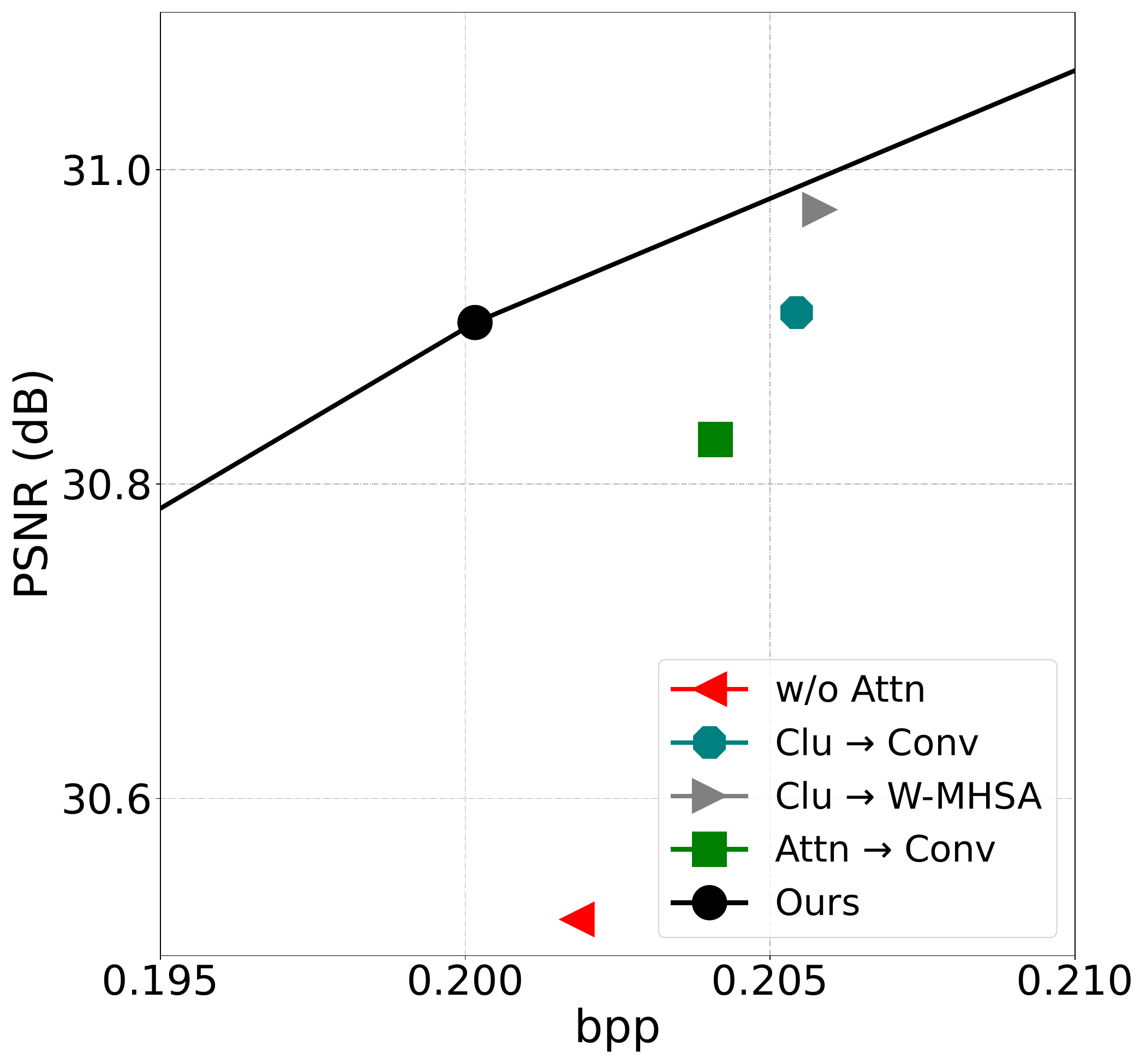}\label{subfig:CCB}} \hspace{\myhspace}
\subfigure[Cluster Variants]{\includegraphics[width=\mywidth\linewidth]{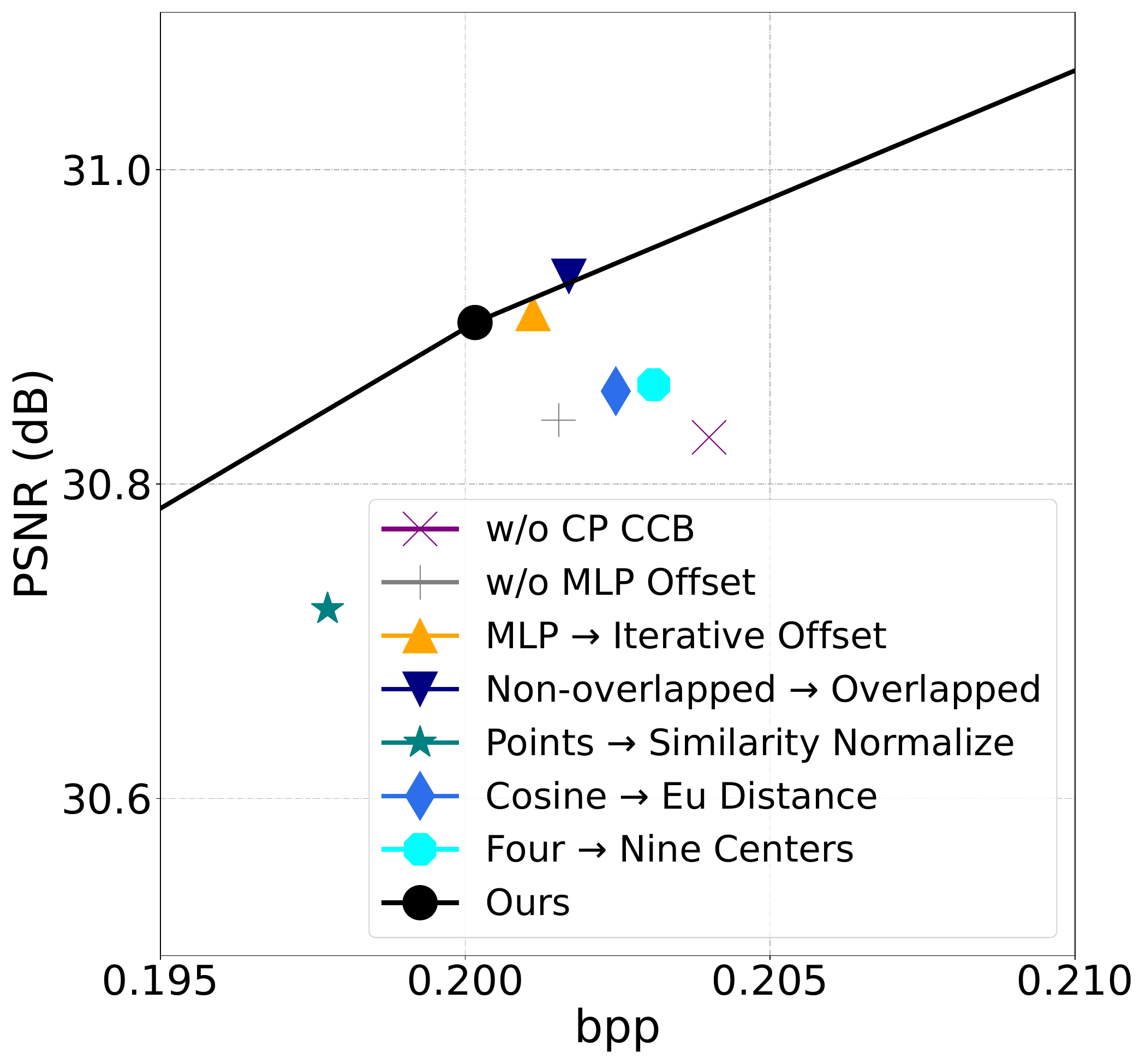}\label{subfig:cluster}} \hspace{\myhspace}
\subfigure[GuidedPQF]{\includegraphics[width=\mywidth\linewidth]{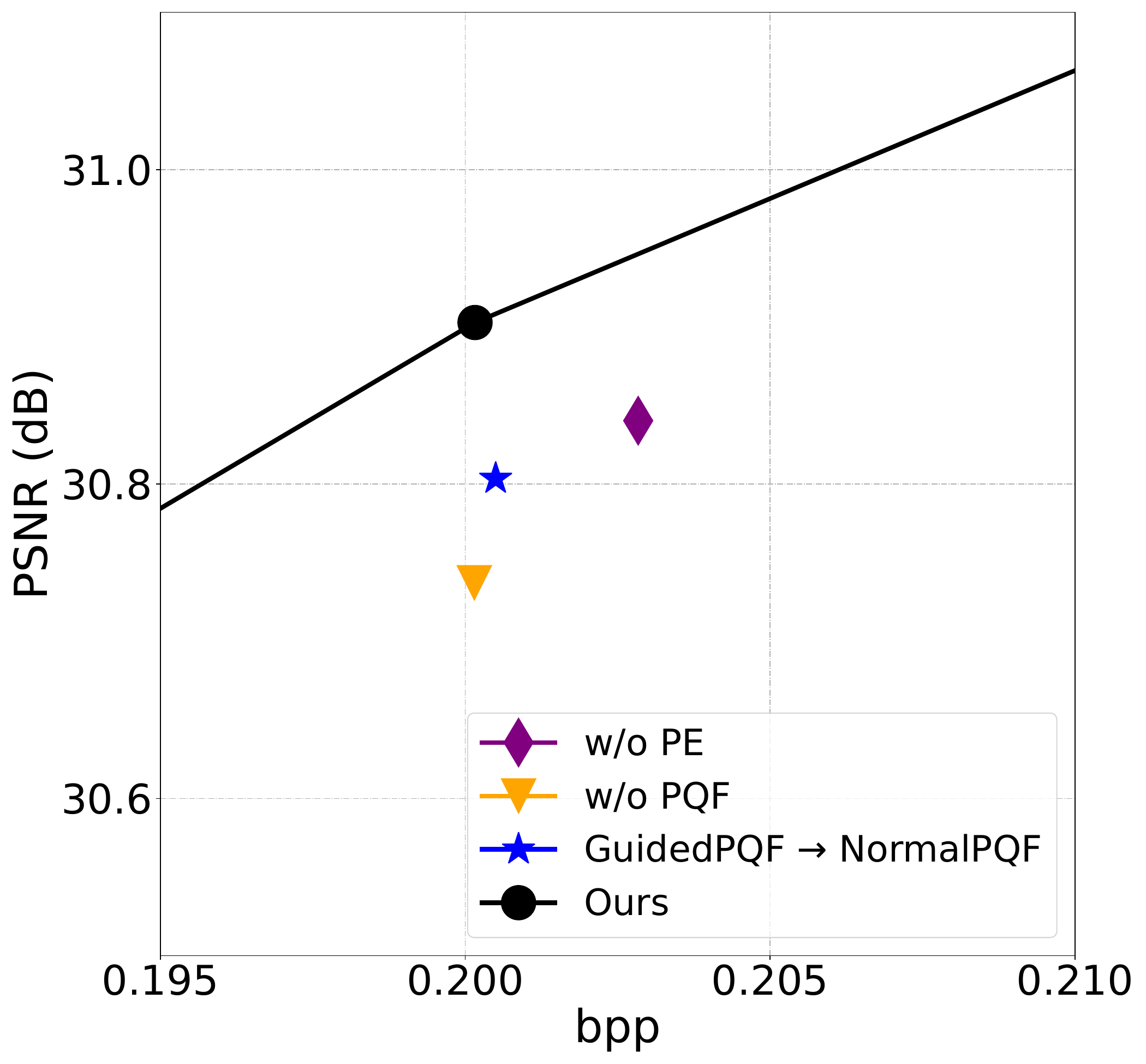}\label{subfig:PQF}} 
\caption{Ablation experiments on modular components.} 
\label{fig:ablation} 
\end{figure*}

\subsection{Ablation Study}
\label{sect:exp_ablation}
A series of ablation studies are conducted to verify the contribution of each modular component in CLIC.

{\bf Positional Encoding.}
Positional encoding (PE) plays a crucial role in CLIC. Figure~\ref{subfig:PQF} shows that ``w/o PE'' suffers from performance degradation. This occurs because pixel relationships are highly correlated with their positions: closer points exhibit higher correlation. Thus, we embed the PE module into CLIC for position-dependent processing.

{\bf Clustering in CCB.}
As illustrated in Figure~\ref{fig:CCB}, the clustering module (Clu for short) acts as a ``Token Mixer'' in CCB. We then implement the clustering module with other token mixers such as conventional CNN or Transformer for the same purpose. Specifically, we employ a 3$\times$3 convolution layer or a Window-based Multi-Head Self-Attention (W-MHSA)~\cite{liu2021swin} to replace the clustering module and remain other modules unchanged. Results reported in Figure~\ref{subfig:CCB} and Table~\ref{tab:tokenmixer} show that the use of W-MHSA causes a 5.14\% increase in model parameters and a 1.94\% reduction in MACs, whereas the use of convolutions increases 12.34\% parameters and 25.02\% MACs.
Although the BD-Rate performance of using W-MHSA is comparable to ours, it greatly increases the encoding and decoding time by 67.35\% and 127.49\%. Due to the efficiency of the convolution-based implementation, its encoding/decoding time is reduced by only 3.49\%/11.33\%; however, its RD performance suffers.
 
{\bf Checkboard pattern in CCB.} Besides, we apply a checkboard pattern at even-numbered CCBs. When this technique is removed, the coding efficiency dramatically suffers, shown as ``w/o CP CCB'' in Figure~\ref{subfig:cluster}.

{\bf Attention in CCB.}
In each CCB, we introduce Spatial Attention and Channel Attention to augment the spatial and channel interaction on top of clustering features. The removal of these two attention units (w/o Attn) leads to significant performance degradation, as shown in~Figure~\ref{subfig:CCB}. In addition, when we replace both attention units with convolutions (see Table~\ref{tab:tokenmixer} and Figure~\ref{subfig:CCB}), the number of parameters, MACs, and coding/decoding time increase significantly, and the performance decreases dramatically. These results further demonstrate the validity of the proposed CCB, where the attention mechanism is effectively and efficiently combined with contextual clustering. 

\begin{table}[t]
  \centering
  
  \fontsize{34pt}{40pt}\selectfont
  \resizebox{\linewidth}{!}   {
    \begin{tabular}{c|c|c|c|c}\hline\hline
          & {$\Delta$ Parameters } & $\Delta$ MACs & $\Delta$ Enc. & $\Delta$ Dec. \\\hline
    Clu $\rightarrow$ Conv  & {12.34\%} & {25.02\%} & -3.49\% & -11.33\% \\
    Clu $\rightarrow$ W-MHSA & {5.14\%} & {-1.94\%}&67.35\%
 &127.49\%
 \\
    CA \& SA $\rightarrow$ Conv &101.27\% &59.93\% & 6.32\%&1.91\% \\
   Ours  & 0\%     & 0\%&0\%&0\% \\\hline
    \end{tabular}}\selectfont%
    \caption{Ablation experiments on Clustering in CCB}
  \label{tab:tokenmixer}%
\end{table}%

{\bf Clustering Variants.}
Next, we conduct ablation studies to discuss clustering-related methods. All results are provided in Figure~\ref{subfig:cluster}.

A point worth exploring is overlapped versus non-overlapped clustering. In CLIC, each pixel point belongs to only one cluster. In addition, we use an overlapped clustering method where each point belongs to two clusters. Figure~\ref{subfig:cluster} shows that ``Non-Overlapped $\rightarrow$ Overlapped'' has little effect on the results while requiring extra complexity. 

Besides, clustering centers also impact the results. We first examine the center offsets. It is observed that using MLP for center offsets yields comparable results to iteratively updating the centers (ten iterations), while ``w/o MLP Offset'' significantly degrades performance. As for the number of clustering centers, using nine clustering centers even performs worse than using four centers, which is probably because the involvement of more clusters makes the divisions too fine, isolating relevant points.

When clustering, to control the magnitude of $F$, we use ${1+m}$ to scale it for normalization. Here, we implement another similarity-based normalization ~\cite{ma2023image}, i.e., $1+\sum_{i=1}^m \operatorname{sigmoid}\left(\alpha \cdot s_i +\beta\right)$. Figure~\ref{subfig:cluster} demonstrates that the similarity-based normalization (Points → Similarity Normalize) yields inferior performance. 

In addition, we use the Euclidean distance ($\text{Eu}$) to evaluate the similarity for clustering: each point is assigned to a center that has the minimum Euclidean distance to it. However, this method yields poor results because the Euclidean similarity only captures magnitude correlation while neglecting orientation and distribution similarity, thus limiting the correlation exploration across latent features.

{\bf GuidedPQF.}
We remove the proposed GuidedPQF and the corresponding coding performance is colored in orange in Figure~\ref{subfig:PQF}. Moreover, we replace the GuidedPQF using a normal MSE-based PQF~\cite{fu2023asymmetric}, called NormalPQF in Figure~\ref{subfig:PQF}. Clearly, both NormalPQF and GuidedPQF attain gains, and GuidedPQF gains more due to the guidance of the additional side information $a_i^j$s.

\section{Conclusion}
\label{sect:conclusion}

In this paper, we present a Contextual Clustering based Learned Image Coding. We start by considering an image as a collection of individual pixel points and systematically reorganize all the points into several clusters via a clustering algorithm to exploit intra-cluster correlations. Then, we reorder all obtained point features according to their positions and further apply local attention mechanisms to the reordered features for inter-cluster correlation exploration. In this way, we achieve global feature characterization of an image, resulting in a more compact representation compared to conventional rectangular shape-based convolutions. 
Extensive experiments validate the superiority of our proposed method over state-of-the-art rule-based VVC and learned image compression methods.

\section*{Acknowledgements}
This work was supported by the National Natural Science Foundation of China (No. U20A20184 and 62171174).

\bibliography{./E2E.bib}


\end{document}